\documentclass[preprint,12pt,showkeys,nofootinbib,floatfix]{revtex4}
\usepackage{amssymb}
\usepackage{amsfonts}
\usepackage{amsmath}
\usepackage{graphicx}
\usepackage{subfigure}
\usepackage{tikz}
\usepackage[
            pdfstartview=FitH,
            bookmarksnumbered=true,
            bookmarksopen=true,
            colorlinks,
            pdfborder=001,
            linkcolor=green,
            anchorcolor=green,
            citecolor=red
            ]{hyperref}
\usepackage[toc,page,title,titletoc,header]{appendix}
\usepackage{graphics}
\usepackage{epsfig}
\usepackage{slashed}
\usepackage{latexsym}
\usepackage{syntonly}
\newcommand{\be}{\begin{equation}}
\newcommand{\ee}{\end{equation}}
\newcommand{\bea}{\begin{eqnarray}}
\newcommand{\eea}{\end{eqnarray}}

\begin{document}
\title{Holographic complexity: A tool to probe the property of reduced fidelity susceptibility}
\author{Wen-Cong Gan$^{1,2}$, Fu-Wen Shu$^{1,2}$}
\thanks{E-mail address:shufuwen@ncu.edu.cn}
\affiliation{
$^{1}$Department of Physics, Nanchang University, Nanchang, 330031, China\\
$^{2}$Center for Relativistic Astrophysics and High Energy Physics, Nanchang University, Nanchang 330031, China}
\begin{abstract}
Quantum information theory along with holography play central roles in our understanding of quantum gravity. Exploring their connections will lead to profound impacts on our understanding of the modern physics and is thus a key challenge for present theory and experiments. In this paper, we investigate a recent conjectured connection between reduced fidelity susceptibility and holographic complexity (the RFS/HC duality for short). We give a quantitative proof of the duality by performing both holographic and field theoretical computations. In addition,  holographic complexity in $AdS_{2+1}$ are explored and several important properties are obtained. These properties allow us, via the RFS/HC duality, to obtain a set of remarkable identities of the reduced fidelity susceptibility, which may have significant implications for our understanding of the reduced fidelity susceptibility. Moreover, utilizing these properties and the recent proposed diagnostic tool based on the fidelity susceptibility, experimental verification of the RFS/HC duality becomes possible.
\end{abstract}
\maketitle

\section{Introduction}

	Recent progress in quantum information theory has shed new light on the deep understanding of quantum gravity. Indeed, many fascinated viewpoints in quantum information theory have been absorbed in the studies of AdS/CFT correspondence\cite{maldacena1,gkp,witten}. One of the most outstanding examples is the discovery that the holographic entanglement entropy--- the area of the minimal surface or Ryu-Takayanagi(RT) surface\cite{RT}---is identified with the entanglement entropy in CFTs. The tensor-network interpratation\cite{swingle} of the RT formula provides further evidence. Especially, it was argued that quantum entanglement is fundamental in building geometries of holographic spacetime\cite{raamsdonk,takayanagi,gsw}. However, just as pointed out by Susskind \cite{susskind}, ``entanglement is not enough", and it is natural to find other quantum information quantities which can be used in studying holography.

One recent achievement is a conjectured duality between quantum complexity \cite{osborne} and the spatial volume of the Einstein-Rosen bridge in AdS\cite{susskind,susskind2} (see e.g.\cite{cai,lu} for recent discussion). Inspired by this conjecture, there is a newly proposed duality\cite{takaya} which connects fidelity susceptibility (also called information metric)\footnote{In the past few years, fidelity has gained its importance in quantum information in studying critical phenomenon (see e.g. \cite{gu} for a review).} and the max volume of a codimension one time slice in the AdS. As a natural generalization, subsystem $A$ of a CFT was considered in\cite{Alish}, where \textit{holographic complexity}(HC) is defined
\begin{equation}\label{holocomplex}
\mathcal C_A=\frac{V(\gamma)}{8\pi RG},
\end{equation}
where $\gamma$ is the RT surface of $A$, $V(\gamma)$ is the volume of the part in the bulk geometry enclosed by the RT surface, and $R$ is the AdS radius (see e.g.\cite{Holographic complexity1,Holographic complexity2,Holographic complexity3,Holographic complexity4} for recent discussion on holographic complexity).
They also found, qualitatively, some similar behaviors between reduced fidelity susceptibility(RFS) and HC. However, quantitative calculation of those behaviors is still missing. One of the main purpose of this work is to  verify, quantitatively, the duality between RFS and the HC. Our results confirm that \textit{the RFS of a subregion $A$ of a CFT is proportional to the volume of the codimension one surface which is enclosed by the subregion $A$ and the RT surface in the dual AdS spacetime, } or by short RFS/HC duality\footnote{During the preparation of our paper, we noticed a related article \cite{BES} appearing on the arXiv. However, in the present paper we focus on the exact marginal perturbation of CFT on the boundary, instead, they considered bulk mass perturbation.}.

The discovered connection between RFS and HC has profound implications for our understanding of quantum many-body systems and quantum information theory. First of all, many properties of the RFS can be easily obtained by using the HC. For example, the Gauss-Bonnet theorem in two dimensional hyperbolic space implies rigorous identities for RFS as we will see below. To the best of our knowledge, these identities are never obtained in the previous literatures and they provide new insights into the nature of the RFS. Ever more, in virtue of these properties, experimental verification of the validity of the RFS/HC duality becomes possible. Below, we address these properties in (2+1) AdS, and their theoretical and experimental implications will also be discussed.

This paper is organized as follows. In section 2, we give an overview of the HC and some of nontrivial properties of the HC in one (spatial) dimension are obtained. Then in section 3, we perform field theoretical computations of the RFS so as to quantitatively confirm the duality between HC and RFS. In section 4, we will try to give a holographic verification of our proposal based on \cite{takaya,bak}. Some significant identities for the RFS are obtained and the potential experimental implications are discussed in section 5. Finally in section 6 we present our main conclusions.

\section{Holographic complexity: an overview}
\label{HC}

Holographic complexity proposed in \cite{Alish} is given by \eqref{holocomplex}. If we consider $A$ to be a ball-shaped region of radius $l$, it is convenient to use the following AdS$_{d+2}$ metric
\begin{equation}
ds^2=\frac{R^2}{r^2}(-dt^2+dr^2+d\rho^2+\rho^2 d\Omega_{d-1}^2),
\end{equation}
then RT surface is parameterized by $\rho=f(r)= \sqrt{l^2-r^2}$ \cite{RT}, and $V(\gamma)$ is given by \cite{Alish}
\begin{align} \label{int}
V&=\Omega_{d-1}R^{d+1} \int_{\rho \le f(r)} d\rho  dr \frac{\rho^{d-1}}{r^{d+1}} \nonumber\\
&=\frac{\Omega_{d-1}R^{d+1}}{d} \int_{\epsilon}^l dr \frac{(l^2-r^2)^{d/2}}{r^{d+1}},
\end{align}
where $\epsilon$ is the UV cutoff.

It is straightforward to perform the integration over $r$ and to obtain the HC. Explicitly, it is
\be\label{odd}
V=\frac{\Omega_{d-1}R^{d+1}}{d}\left(\frac{l^d}{d\epsilon^d}-\frac{d}{2(d-2)}\frac{l^{d-2}}{\epsilon^{d-2}}+\frac{d(d-2)}{8(d-4)}\frac{l^{d-4}}{\epsilon^{d-4}}+\dots-(-1)^{[\frac d2]}\frac{\pi}{2}\right),
\ee
for odd $d$, and
\be
V=\frac{(-1)^{\frac d 2}\Omega_{d-1}R^{d+1}}{d} \left[ \ln\left (\frac l \epsilon\right) +\sum_{k=1}^{\frac d 2} {k\choose \frac d 2} \frac{(-1)^{k}}{2k} \left(\frac l \epsilon\right)^{2k}+\sum_{k=1}^{\frac d 2} {k\choose\frac d 2}\frac{(-1)^{k+1}}{2k}\right] \label{even}
\ee
for even $d$.

In this paper, we pay our attention to the $d=1$ case. In this case the ball-shaped region becomes an interval, and \eqref{odd} implies that the volume bounded by the RT surface is given by
\begin{equation} \label{v}
V_A=2R^{2} \left[\frac l \epsilon -\frac \pi 2\right].
\end{equation}
The HC $\mathcal C_A$ defined in \eqref{holocomplex} is then given by \cite{Alish}\footnote{We find that there is a factor 2 missing in the equation(19) of \cite{Alish}, for $l$ is the radius of the region to be concerned.}
\begin{equation} \label{hc}
\mathcal C_A=\frac{cl}{6\pi\epsilon}-\frac{c}{12},
\end{equation}
where $c=\frac{3R}{2G}$ is the central charge of the CFT, and $l$ is the half of the length of the interval. In what following, we would like to show that there are many interesting properties of the holographic complexity, which are of particular significance in our understanding of the fidelity susceptibility.

\subsection{Property 1}
\begin{figure}[h]
\begin{tikzpicture}[scale=2.5]
\draw (-1.5,0) -- (1.5,0);

\filldraw[fill=gray!40] (1cm,0cm) arc (0:180:1cm) node[anchor=north]{$x_1$};
\filldraw[fill=white] (0.2cm,0cm)node[anchor=north]{$x_2$} arc (0:180:0.6cm);
\filldraw[fill=white] (1cm,0cm)node[anchor=north]{$x_3$} arc (0:180:0.4cm);

\end{tikzpicture}
\caption{Two subregions $A$ and $B$ which share an endpoint. Shaded part denotes the hyperbolic triangle.}
\label{fig1}
\end{figure}
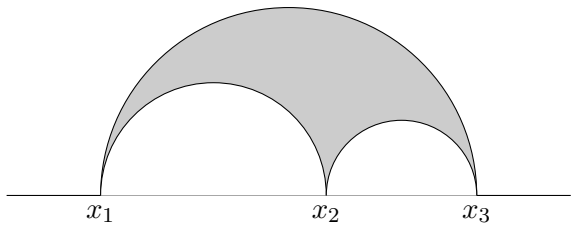

For two given subregions $A$ and $B$, if $A\cap B = \varnothing$ except they share an endpoint (see Fig.\ref{fig1}), then $\mathcal C_{A\cup B}-\mathcal C_A -\mathcal C_B = constant$. Constant means that it depends neither on coordinate nor on the scale of the region.

The above argument is very easily to prove. Let $A$ be the interval $\{x_1, x_2\}$, and $B$ be the interval $\{x_2, x_3\}$, then $A\cup B$ is $\{x_1, x_3\}$ (see Fig.\ref{fig1}). The holographic complexity \eqref{hc} then satisfies
\begin{equation}\label{prop1}
\mathcal C_{A\cup B}-\mathcal C_A -\mathcal C_B =\frac{c}{12},
\end{equation}
which is independent of the coordinate and the scale of the region, and is proportional to the central charge.

Actually, this is nothing but the Gauss-Bonnet theorem in hyperbolic space if we notice that  $V_{A\cup B}-V_A -V_B$ is the volume of a hyperbolic triangle (the shaded region in Fig.\ref{fig1}).
The Gauss-Bonnet theorem states that the area of a hyperbolic triangle with internal angles $\alpha$, $\beta$, $\gamma$ is equal to $\pi-(\alpha+ \beta+ \gamma)$ with $R$ set to 1\footnote{By applying the M\"{o}bius transformation one can always set $R$ to $1$.}. In the case of Fig.\ref{fig1}, $\alpha= \beta= \gamma=0$, we then have
\begin{equation}
\label{eq:x}
V_{A\cup B}-V_A -V_B =\pi.
\end{equation}
Back to the holographic complexity, it reduces to \eqref{prop1} (with $R=1$).

One point which is worthy of mention is that \eqref{prop1} depends only on the central charge. From symmetry point of view, this implies that it is invariant under conformal transformation. Indeed, this is consistent with the property of hyperbolic ideal triangle which area of hyperbolic ideal triangle under M\"obius transformation remains unchanged due to the Gauss-Bonnet theorem.

\begin{figure}[h]
\begin{tikzpicture}[scale=3]
\draw (-1.5,0) -- (1.5,0);
\fill [fill=gray!40](0.6cm,0cm)node[anchor=north]{$l$} arc (0:180:0.6cm) node[anchor=north]{$-l$}(-0.6,0)--(-0.6,1)--(0.6,1)--(0.6,0);
\filldraw (0,0) circle (0.01)node[anchor=north]{$0$};
\draw (0.6cm,0cm)node[anchor=north]{$l$} arc (0:180:0.6cm) node[anchor=north]{$-l$};
\draw (0.6,0)--(0.6,1);
\draw (-0.6,0)--(-0.6,1);
\end{tikzpicture}
\caption{A hyperbolic triangle with infinite endpoints}
\label{fig1-infty}
\end{figure}
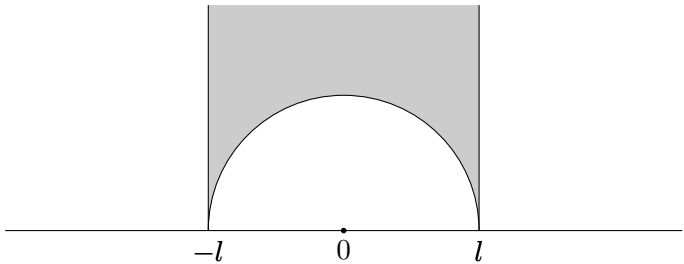

This result can be extended to the case where the hyperbolic triangle includes $\infty$ as its endpoints (as shown in Fig. \ref{fig1-infty}).

In this case, the HC of the hyperbolic triangle is also a constant. To see this, we notice that the AdS$_3$ metric is given by
\begin{align}
ds^2=\frac{R^2}{r^2}(-dt^2+dr^2+d\rho^2).
\end{align}
As $r$ approaches infinity, the distance between points $(r,\rho_1)$ and $(r, \rho_2)$ approaches $0$, and the two infinity endpoints can be viewed as one same point.

The volume and the HC of the infinite rectangle region are given by
\begin{align}
V&=\int \frac{R^2}{r^2} dr d\rho= R^2 \int_{-l}^{l}d\rho \int_{\epsilon}^{\infty}\frac{dr}{r^2}=\frac{2R^2 l}{\epsilon}, \\
\mathcal C&=\frac{V}{8\pi RG}=\frac{cl}{6\pi \epsilon}.
\end{align}
As a result, the holographic complexity of the hyperbolic triangle is given by
\begin{align}
\mathcal C_{\triangle}=\mathcal C- \mathcal C_A=\frac{c}{12},
\end{align}
which is the same as the one given in \eqref{prop1}.

\subsection{Property 2}
\begin{figure}[h]
\begin{tikzpicture}[scale=2.2]
\draw (-1.5,0) -- (1.5,0);

\filldraw[fill=gray!40] (1.4cm,0cm) arc (0:180:1.4cm) node[anchor=north]{$x_1$};
\filldraw[fill=white] (-1cm,0cm)node[anchor=north]{$x_2$} arc (0:180:0.2cm);
\filldraw[fill=white] (-0.8cm,0cm)node[anchor=north]{$x_3$} arc (0:180:0.1cm);
\filldraw[fill=white] (-0.5cm,0cm)node[anchor=north]{$x_4$} arc (0:180:0.15cm);

\filldraw[fill=white] (1.4cm,0cm)node[anchor=north]{$x_{n+1}$} arc (0:180:0.3cm) node[anchor=north]{$x_{n}$};
\node   at (-0.2,0.05) {$\cdots$};
\end{tikzpicture}
\caption{Geodesics of $n$ subregions $A_i$ and their union $\bigcup  \limits_{i}A_i$ enclose an $(n+1)$-sided hyperbolic polygon as shown in the shaded region.}
\label{fig2}
\end{figure}
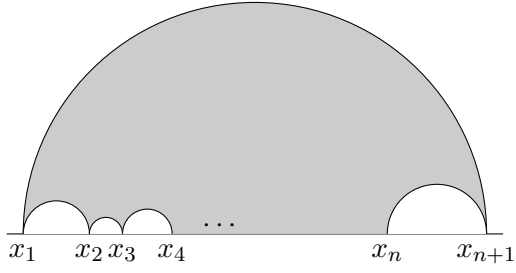

One direct generalization of the above property is the case with arbitrary $n$($n\geq 2$) intervals. The difference between the total HC and the sum of individual HCs is proportional to $n-1$, (see Fig.\ref{fig2}) i.e.
\begin{equation}
\mathcal C_{total}-\sum_i^n \mathcal C_i=\frac{c}{12}(n-1),
\end{equation}
where $\mathcal C_i$ denotes the HC of the interval $\{x_{i}, x_{i+1}\}$, $\mathcal C_{total}$ denotes the HC of the interval $\{x_1, x_{n+1}\}$. The proof is direct as we apply Eq. \eqref{hc}
\begin{equation}
\mathcal C_{total}-\sum_i^n \mathcal C_i =\left[\frac{c(x_{n+1}-x_1)}{12\pi\epsilon}-\frac{c}{12}\right]-\sum_i^n \left[\frac{c(x_{i+1}-x_{i})}{12\pi\epsilon}-\frac{c}{12}\right] =\frac{c}{12}(n-1).
\end{equation}

This is, again, the Gauss-Bonnet theorem for a hyperbolic polygon which states that the volume of an $(n+1)-$sided hyperbolic polygon with vanishing internal angles is proportional to $(n-1)\pi$.
Explicitly, it is (set $R=1$ as before)
\begin{equation}
V_{total}-\sum_i^n V_i =2\left[ \left[\frac {(x_{n+1}-x_1)}{2 \epsilon} -\frac \pi 2\right]-\sum_i^n \left[\frac {(x_{i+1}-x_{i})}{2 \epsilon} -\frac \pi 2\right]\right] =\pi (n-1).
\end{equation}
The result can be also verified by noticing that the shaded region is made of $(n-1)$ ``hyperbolic triangle", such that the volume is $V_A$ multiplied by $(n-1)$ and the HC is $\mathcal C_A$ multiplied by $(n-1)$ too.

\begin{figure}[h]
\begin{tikzpicture}[scale=2.2]
\draw (-1.5,0) -- (1.5,0);

\fill [fill=gray!40](-1.4,0)--(-1.4,1)--(1.4,1)--(1.4,0);
\filldraw[fill=white] (-1cm,0cm)node[anchor=north]{$x_2$} arc (0:180:0.2cm);
\filldraw[fill=white] (-0.8cm,0cm)node[anchor=north]{$x_3$} arc (0:180:0.1cm);
\filldraw[fill=white] (-0.5cm,0cm)node[anchor=north]{$x_4$} arc (0:180:0.15cm);

\filldraw[fill=white] (1.4cm,0cm)node[anchor=north]{$x_{n}$} arc (0:180:0.3cm) node[anchor=north]{$x_{n-1}$};
\node   at (-0.2,0.05) {$\cdots$};
\draw (1.4,0)--(1.4,1);
\draw (-1.4,0)node[anchor=north]{$x_1$}--(-1.4,1);
\end{tikzpicture}
\caption{ $(n+1)$-sided hyperbolic polygon with two endpoints at infinity.}
\label{fig2-infty}
\end{figure}
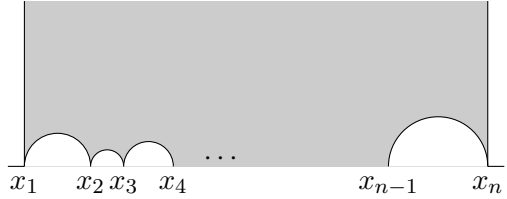

Similarly, one can extend this result to the case where $\infty$ is also an endpoint (as depicted in Fig.\ref{fig2-infty}).  The shaded region in Fig.\ref{fig2-infty} is made of $(n-1)$ pieces of hyperbolic triangles which include infinity as their endpoints, so the volume is equal to $\pi (n-1)$ and the HC is equal to $\frac{c}{12}(n-1)$ as expected.

\subsection{Property 3}
For two given subregions $A$ and $B$, if $A\cap B \neq \varnothing$, then $\mathcal C_{A\cup B}-\mathcal C_A -\mathcal C_B + \mathcal C_{A\cap B}=0$.
\begin{figure}[h]
\centering
  \includegraphics[width=.4\textwidth]{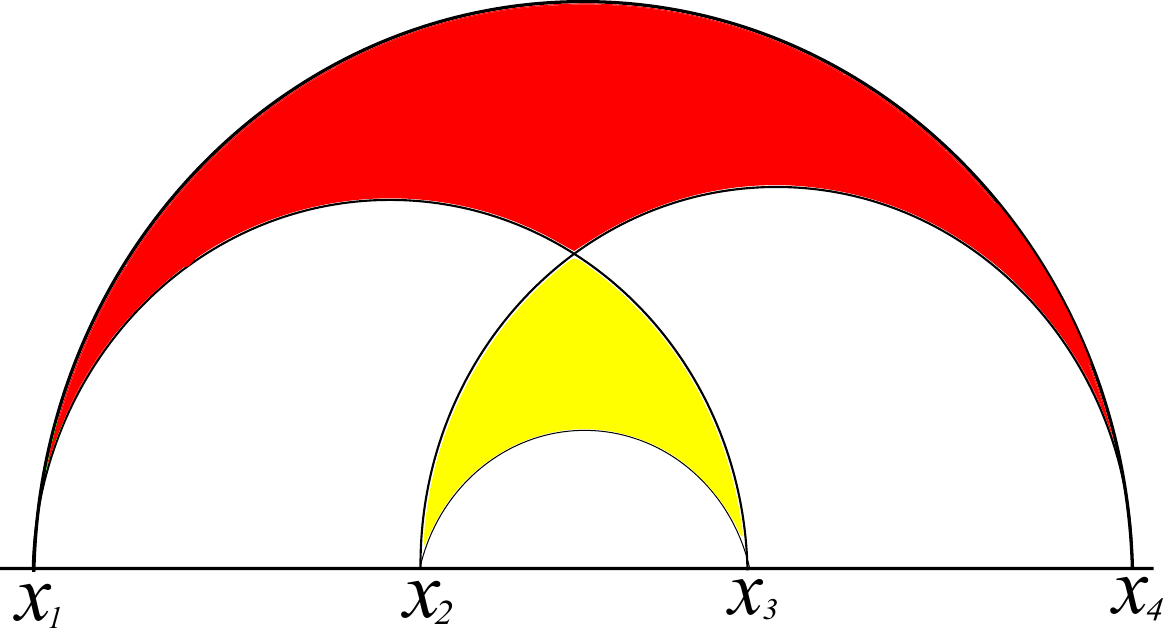}\hspace{-1cm}
\caption{Two subregions with nonvanishing intersection.}
\label{fig3}
\end{figure}

Let $A=\{x_1, x_3\}, B=\{x_2, x_4\}$, then $A\cup B=\{x_1, x_4\}, A\cap B=\{x_2, x_3\}$ as shown in Fig.\ref{fig3}. Substituting these into \eqref{hc} we quickly get
\begin{align}
\mathcal C_{A\cup B}-\mathcal C_A -\mathcal C_B + \mathcal C_{A\cap B} =0.
\end{align}
Actually, this is to say that the red triangle and the yellow triangle (color online) in Fig. \ref{fig3} have the same area. This is consistent with the Gauss-Bonnet theorem in the hyperbolic geometry.

\section{Field theoretical calculation}
In this section, we would like to give an analytical computation of the RFS in more detail, so as to verify the conjectured RFS/HC duality quantitatively. For simplification, we focus on a ball-shaped subsystem of the whole system which is in the vacuum state of a $1+1$ dimensional CFT.

To proceed, let us consider a CFT perturbed by a marginal primary operator. We assume its action can be written as $S=\int d\tau d^d x (\mathcal L+\lambda\cdot O)$, where $\lambda$ is the coupling parameter. Denote its ground state to be $|\psi(\lambda) \rangle $, and the reduced density matrix to be $\rho_{A}(\lambda)$. We want to calculate the RFS between two nearby states $\rho_{A}(\lambda_1) $ and $\rho_{A}(\lambda_2) $, where $\delta\lambda=\lambda_2-\lambda_1$ is small. The reduced fidelity is defined as $F_{A} = \mathrm{Tr} \sqrt{\sqrt{\rho_{A}(\lambda_1)} \rho_{A}(\lambda_2)  \sqrt{\rho_{A}(\lambda_1)}}$ \cite{zhq} and can be expanded as $F_{A} = 1- G_{A} \delta\lambda^2 + \mathcal O(\delta\lambda^3)$, where $G_{A}$ is the fidelity susceptibility \cite{gu}
\begin{equation}\label{fidelity}
G_{A}=\lim_{\delta \lambda \rightarrow 0} \frac{-2 \ln F_{A}}{\delta \lambda^2}=-\frac{\partial^2 F_{A}}{\partial (\delta \lambda)^2}.
\end{equation}

We would like to calculate fidelity susceptibility by using the trick introduced by Casini et al. in \cite{CHM}, namely, we conformally map the domain of dependence of a ball-shaped region $A \subset \mathbb R^{d,1}$ to a hyperbolic cylinder $\mathbb H_{d} \times \mathbb R$. The map starts with polar coordinates in flat space
\begin{equation}\label{metric}
ds^2=-dt^2+d\rho^2+\rho^2 d\Omega_{d-1}^2.
\end{equation}
If we introduce the following coordinate transformation
\begin{equation}
\begin{split}
t=l\frac{\sinh(\frac{t'}{l})}{\cosh u +\cosh (\frac{t'}{l})} \,,
\qquad
r=l\frac{\sinh u}{\cosh u +\cosh (\frac{t'}{l})}\,,
\end{split}
\end{equation}
the metric \eqref{metric} then becomes
\begin{equation}
ds^2=\frac{1}{[\cosh u +\cosh (\frac{t'}{l})]^2} \left(-dt'^2+ l^2 (du^2+\sinh^2 u d\Omega_{d-1}^2) \right),
\end{equation}
which is obviously  related to the metric on the  $\mathbb H_d \times \mathbb R$ conformally. Further more, the reduced density matrix $\rho_{A}$ of the subregion $A$ is mapped to a thermal density matrix $\rho_{\mathbb H_d}= e^{-\beta \mathcal H_{\mathbb H_d}}$ on the hyperbolic cylinder $\mathbb H_d \times \mathbb R$ with $\beta=2\pi l$, where $\mathcal H_{\mathbb H_d}$ is the CFT Hamiltonian on $\mathbb H_d \times \mathbb R$, i.e.
\begin{equation}
\rho_{A}= U \rho_{\mathbb H_d} U^\dagger,
\end{equation}
for some unitary operator $U$ which is determined by the geometric conformal map. In Euclidean signature, the map is from $\mathbb R^{d+1}$ to $\mathbb H_{d} \times S^1$.


The key point is that the fidelity calculated on the hyperbolic cylinder
\begin{equation}
F_{\mathbb H_d} = \mathrm{Tr} \sqrt{\sqrt{\rho_{\mathbb H_d}(\lambda_1)} \rho_{\mathbb H_d}(\lambda_2)  \sqrt{\rho_{\mathbb H_d}(\lambda_1)}}
\end{equation}
equals $F_{A} $. The proof is direct if we using the fact that $\sqrt{U\rho U^{\dagger}}=U \sqrt{\rho}U^{\dagger}$ for any unitary matrix $U$. This property allows us to avoid the calculation of $G_{A}$ directly, instead, we calculate the fidelity susceptibility $G_{\mathbb H_d}$ (in hyperbolic cylinder with inverse temperature $\beta=2\pi l$) which is defined through $F_{\mathbb H_d} = 1- G_{\mathbb H_d} \delta\lambda^2 + \mathcal O(\delta\lambda^3)$.


As to the $d=1$ case, the hyperbolic cylinder is $\mathbb H^1 \times S^1$ with the metric
\begin{equation}\label{ds}
ds^2= d\tau^2+l^2 du^2,
\end{equation}
 which is nothing but $\mathbb R^1 \times S^1$. To proceed, we use the trick introduced in \cite{takaya} to compute $G_{\mathbb H^1}$ on $\mathbb R^1 \times S^1$, where fidelity susceptibility can be expressed by
\begin{equation}
G_{\mathbb H^1}=\frac12 \int_{-\infty}^{\infty} dx_1 dx_2 \int_{\frac{3\beta}{8}+\varepsilon}^{\frac{5\beta}{8}-\varepsilon} d\tau_2 \int_{-\frac{3\beta}{8}+\varepsilon}^{\frac{3\beta}{8}-\varepsilon}   d\tau_1  \det(g) \langle O(x_1,\tau_1) O(x_2,\tau_2) \rangle,
\end{equation}
where $\det(g)$ is the determinant of the metric \eqref{ds} and $\langle O(x_1,\tau_1) O(x_2,\tau_2) \rangle$ is the two point function on $\mathbb R^1 \times S^1$,
\begin{equation}
\langle O(x_1,\tau_1) O(x_2,\tau_2) \rangle = \frac{(\frac\pi\beta)^{2\Delta}}{ \left(\sinh^2 (\frac{\pi(x_1-x_2)}{\beta})+\sin^2 (\frac{\pi(\tau_1-\tau_2)}{\beta})\right)^\Delta},
\end{equation}
where $\Delta$ is the conformal dimensions. In the present work, we focus on the case where the  perturbation is exactly marginal $\Delta=2$. After using the thermal periodicity $\beta=2\pi l$, we finally get
\begin{equation}
G_{\mathbb H^1}=\frac{V_1 l}{8}\left(\frac{\pi l}{\varepsilon}-\frac12\right).
\end{equation}
Comparing this with $F_{\mathbb H_d}=1 - 2 \pi l \chi_ \lambda \frac{\delta\lambda^2}{8}+ \mathcal O (\delta\lambda^3)$ \cite{Alish} yields the following result
\begin{equation}
\chi_ \lambda=\frac{V_1 }{2\pi}\left(\frac{\pi l}{\varepsilon}-\frac12\right),
\end{equation}
where $\chi_{\lambda}$ is the RFS.
If we set $\varepsilon=\pi^2 \epsilon$, and let the IR regulator $V_1$ equals to $\frac{\pi c}{3}$, then
\begin{equation}
\chi_ \lambda = \frac{cl}{6\pi\epsilon}-\frac{c}{12} =\mathcal C_A.
\end{equation}
The above calculations quantitatively show that the RFS/HC duality rigorously holds at least in (1+1) CFT with marginal perturbations.

\section{Holographic calculation: more evidences}
Now we turn to the holographic calculation. We will follow the method in \cite{takaya,bak} and review it first. The CFT perturbed by a primary operator $O$ is dual to the Einstein-scalar system with the action given by
\begin{equation}\label{action}
I=-\frac{1}{16\pi G} \int d^{d+2}x \sqrt g \left[\mathcal R- g^{ab}\partial_a \phi \partial_b \phi+ \frac{d(d+1)}{R^2}\right],
\end{equation}
where $\phi$ is the bulk scalar field dual to primary operator $O$, $R$ is the AdS radius.

In three dimensions, the solution to the Einstein equation is the so-called Janus solution\cite{Janus1,Janus2}
\begin{align}
ds^2&=R^2[dy^2+ f(y)ds^2_{AdS_2}], \label{janus}\\
f(y)&=\frac 1 2 (1+\sqrt{1-2\alpha^2}\cosh (2y)), \\
\phi(y)&=\phi_0+ \alpha \int_{-\infty}^y \frac{dy}{f(y)}.
\end{align}
with $\delta\lambda=\lambda_2-\lambda_1=2\alpha+\mathcal O(\alpha^3)$.

The metric \eqref{janus} should match the undeformed pure AdS metric $ds^2_{pure}=R^2\left(d \hat y^2+ \frac 1 2 (1+\cosh (2\hat y))ds^2_{AdS_2}\right)$ at the infinity $|y|=y_{\infty} \rightarrow \infty$. This leads to the condition
\begin{equation}
\sqrt{1-2\alpha^2} e^{2y_\infty}=e^{2\hat y_\infty}.
\end{equation}

Next let us focus on the ball-shaped subregion $A$ on the time slice of the CFT (which is an interval in the case of CFT$_{1+1}$). That is to say, we should concentrate on the on-shell action \eqref{action} of the region bounded by the RT surface $\gamma$ of A. This is evaluated by
\begin{equation}
I_\alpha=\frac{1}{4\pi R G} V(\gamma) \int_{-y_\infty}^{y_\infty} dy f(y),
\end{equation}
where $V(\gamma)$ is the volume of the part in the bulk geometry enclosed by the RT surface and is given by \eqref{int} of the $d=1$ case. The difference of the on-shell action is given by
\begin{equation}
I_\alpha-I_0= \frac{1}{16\pi R G} V(\gamma) \log (\frac{1}{1-2\alpha^2}).
\end{equation}
As in \cite{takaya,bak}, fidelity can be calculated as $F_{\mathbb H^1}=e^{-(I_\alpha-I_0)}$. If $\alpha$ is small, we have
\begin{equation}
F_{\mathbb H^1}=e^{-(I_\alpha-I_0)} \simeq 1- \frac{1}{8\pi R G} V(\gamma) \alpha^2= 1- \frac{1}{32\pi R G} V(\gamma) (\delta \lambda)^2,
\end{equation}
which implies that the fidelity susceptibility is given by
\begin{equation}
G_{\mathbb H^1}=\frac{1}{32\pi R G} V(\gamma).
\end{equation}

For higher dimensional case, as demonstrated in \cite{takaya}, the holographic dual of marginal deformation CFT is approximated by adding  a defect brane $\Sigma$ with a tension $T$ in the AdS spacetime, extending from AdS boundary to the bulk and locate on the relevant time slice. The defect brane action is
\begin{equation}
I_{brane}=T\int_{\Sigma}\sqrt g.
\end{equation}
Once we focus on the subregion, the region bounded by RT surface is to be considered, and the defect brane action in this region is given by $\tilde I_{brane}=T\int_{\gamma}\sqrt g$ where the subscript $\gamma$ means the integral is over the region bounded by RT surface $\gamma$.
The deformed action can be written as $I_\alpha=I_0+\tilde I_{brane}$. By making use of $F_{\mathbb H^1}=e^{-(I_\alpha-I_0)}$, we have
\begin{equation}
F_{\mathbb H^1}=e^{-(I_\alpha-I_0)}=e^{-\tilde I_{brane}} \simeq 1- T\int_{\gamma}\sqrt g.
\end{equation}
The fidelity susceptibility turns out to be
\begin{equation}
G_{\mathbb H^1}\propto \int_{\gamma}\sqrt g= V(\gamma).
\end{equation}
In this way, the RFS/HC duality in higher dimensions is also confirmed.

\section{Properties of the RFS and their experimental implications}
In the previous sections, we have shown explicitly that the HC is dual to the RFS. In this section, we would like to address more detail about its potential significance and some possible experimental implications.

Given the duality between HC and RFS, properties of the HC which are addressed in section \ref{HC} immediately imply that the same properties should be hold for the RFS in ($1+1$) CFT. More specifically, the following identities should be hold for the RFS:

(i) For two given subsystems $A$ and $B$, if $A\cap B = \varnothing$ but they are adjacent to each other, then
\bea
\chi_{A\cup B}-\chi_A -\chi_B =\frac{c}{12}. \label{three}
\eea
(ii) Generalization to the case with arbitrary $n$($n\geq 2$) adjacent subsystems yields
\begin{equation}
\chi_{total}-\sum_i^n \chi_i=\frac{c}{12}(n-1).\label{n}
\end{equation}
(iii) For two given subsystems $A$ and $B$, if $A\cap B \neq \varnothing$, then
\be
\chi_{A\cup B}-\chi_A -\chi_B + \chi_{A\cap B}=0.\label{union}
\ee

These identities are never obtained before. The conceptual importance of these identities is huge.

First of all, they make the experimental verification of the RFS/HC duality possible. Recent theoretical and experimental studies show that fidelity susceptibility is related to the dynamical response of the many-body system which is measurable using spectroscopy measurements \cite{hhtz}, or using time-dependent quenches or ramps\cite{KGP}. In addition, it is also suggested that fidelity susceptibility can be obtained from the overlap of quantum wave functions, which is detectable in an NMR quantum simulator \cite{jzhang} and in ultracold bosons by applying a quantum gas microscope\cite{islam}. Therefore, in principle, with these detection approaches we can verify the validity of the RFS/HC duality by  examining \eqref{three}-\eqref{union} experimentally.

Second, these identities have fundamental theoretical implications. We can see all formulae \eqref{three}-\eqref{union} rely neither on coordinate nor on the scale of the region, it is thus a reflection of the conformal symmetry of the system which is consistent with the Gauss-Bonnet theorem. Furthermore the fact that \eqref{three} and \eqref{n} just depend on the central charge means that the HC of the hyperbolic ideal triangle is a kind of measure of the degrees of freedom of the system and may provide a new tool to investigate the holographic c-theorem. In this sense, besides entanglement entropy, RFS can be a new CFT quantity to probe bulk under reorganization and be significant in bulk reconstruction. Thus it is an interesting future problem to investigate the relation between RFS and quantum error correction \cite{error,error2}.


\section{Conclusions}
In this paper, we have given an explicit proof of the recent conjectured connection between RFS and HC, which we call it the RFS/HC duality. Our verification includes two parts. Firstly we perform geometric calculations and obtain the explicit expression of the HC for arbitrary dimensions. We then focus on the field theoretical computations of the RFS in $(1+1)$ CFT. In this way we find the expressions of the RFS and of the HC are exactly same, at least for $(1+1)$ CFT. To find more evidences especially for higher dimensions, holographic calculations were also performed, and as expected, we reach the same claim, i.e., the RFS/HC duality is valid even for higher dimensions.

As a second achievement of this paper, we also investigated several important properties of the HC in $AdS_{2+1}$. These properties lead to, via the RFS/HC duality, a set of remarkable identities of the RFS, which may have fundamental implications for our understanding of the RFS both in theoretical and experimental aspects. Moreover, utilizing these properties and the recent proposed diagnostic tool based on the fidelity susceptibility, experimental verification of the RFS/HC duality becomes possible.

In order to better understand the deep meaning of the identities of the RFS, and give a further verification of these identities, it is worthwhile for future studies to explore several models (e.g., the Ising model) in more detail. In addition, it is also very interesting to perform the field theoretical computations in higher dimensions in the future studies. As an another future direction, notice that ground state fidelity can also be expressed via tensor network \cite{vidal}. Therefore, it is very interesting to investigate the duality between reduced fidelity susceptibility and holographic complexity using tensor network.

\begin{acknowledgments}
This work was supported in part by the National Natural Science Foundation of China under Grant No. 11465012, the Natural Science Foundation of Jiangxi Province under Grant No. 20142BAB202007 and the 555 talent project of Jiangxi Province.
\end{acknowledgments}


\begin{thebibliography}{99}


\bibitem{maldacena1}
J.M. Maldacena, ``The large-N limit of superconformal field theories and supergravity", Adv. Theor. Math. Phys. {\bf2} (1998) 231 [Int. J. Theor. Phys. {\bf38} (1999) 1113] [hep-th/9711200].


\bibitem{gkp}S. S. Gubser, I. R. Klebanov and A. M. Polyakov, ``Gauge theory correlators from non-critical string theory," Phys. Lett. {\bf B} {\bf428} (1998) 105 [hep-th/9802109].

\bibitem{witten}E. Witten, ``Anti-de Sitter space and holography," Adv. Theor. Math. Phys. {\bf2} (1998) 253 [hep-th/9802150].

\bibitem{RT}
S. Ryu, T. Takayanagi, ``Holographic derivation of entanglement entropy from the anti-de sitter space/conformal field theory correspondence", Phys. Rev. Lett. {\bf96} (2006) 181602 [arXiv:hep-th/0603001].

\bibitem{swingle}
B. Swingle, ``Entanglement Renormalization and Holography", Phys. Rev. {\bf D} {\bf 86} (2012) 065007 [arXiv:0905.1317[cond-mat]].

\bibitem{raamsdonk}
M. Van Raamsdonk, ``Building up spacetime with quantum entanglement", Gen. Rel. Grav. {\bf 42}
(2010) 2323 [Int. J. Mod. Phys. {\bf D} {\bf 19} (2010) 2429] [arXiv:1005.3035].

\bibitem{takayanagi}
M. Nozaki, S. Ryu and T. Takayanagi, ``Holographic geometry of entanglement renormalization
in quantum field theories", JHEP {\bf 10} (2012) 193 [arXiv:1208.3469].

\bibitem{gsw}
W.-C. Gan, F.-W. Shu, M.-H. Wu, ``Thermal geometry from CFT at finite temperature", Phys. Lett. {\bf B} {\bf760} (2016) 796
 [arXiv:1605.05999[hep-th]]; ``Emergent geometry, thermal CFT and surface/state correspondence'', [arXiv: 1606.07628[hep-th]].

\bibitem{osborne}
T. J. Osborne, ``Hamiltonian complexity", Rep. Prog. Phys. {\bf 75}
(2012) 022001 [arXiv:1106.5875].

\bibitem{susskind} L. Susskind, ``Addendum to Computational Complexity and Black Hole Horizons," [arXiv:1403.5695[hep-th]]; ``Entanglement is not enough," [arXiv:1411.0690[hep-th]].

\bibitem{susskind2} A. R. Brown, D. A. Roberts, L. Susskind, B. Swingle, and Y. Zhao, ``Holographic
Complexity Equals Bulk Action?", Phys. Rev. Lett. {\bf 116} (2016) 191301 [arXiv:1509.07876];
A. R. Brown, D. A. Roberts, L. Susskind, B. Swingle, and Y. Zhao, ``Complexity,
action, and black holes", Phys. Rev. {\bf D} {\bf 93} (2016) 086006
[arXiv:1512.04993].

\bibitem{cai}
R.-G. Cai, S.-M. Ruan, S.-J. Wang, R.-Q. Yang and R.-H. Peng, ``Action growth for AdS black holes", JHEP {\bf 1609} (2016) 161 [arXiv: 1606.08307].

\bibitem{lu}
H. Huang, X.-H. Feng and H. L\" u, ``Holographic Complexity and Two Identities of Action Growth", [arXiv:1611.02321].


\bibitem{takaya}
M. Miyaji, T. Numasawa, N. shiba, T. Takayanagi and K. Watanabe, ``Distance between Quantum States and Gauge-Gravity Duality", Phys. Rev. Lett. {\bf115} (2015) 261602 [arXiv:1507.07555[hep-th]].

\bibitem{gu} S-J Gu, ``Fidelity approach to quantum phase transition", Int. J. Mod. Phys. {\bf B} {\bf 24} (2010) 4371 [arXiv:0811.3127[quant-ph]].

\bibitem{Alish}  M. Alishahiha, ``Holographic Complexity", Phys. Rev. {\bf D} {\bf92} (2015) 126009 [arXiv:1509.06614 [hep-th]].

\bibitem{Holographic complexity1}
Davood Momeni, Mir Faizal, Sebastian Bahamonde, Ratbay Myrzakulov, ``Holographic complexity for time-dependent backgrounds", Phys. Lett. {\bf B} {\bf 762} (2016) 276-282.

\bibitem{Holographic complexity2}
N.S. Mazhari, Davood Momeni , Sebastian Bahamonde , Mir Faizal , Ratbay Myrzakulov, ``Holographic Complexity and Fidelity Susceptibility as Holographic Information Dual to Different Volumes in AdS", Phys. Lett. {\bf B} {\bf 766} (2017) 94-101, [arXiv:1609.00250].

\bibitem{Holographic complexity3}
Davood Momeni, Mir Faizal, Kairat Myrzakulov, Ratbay Myrzakulov, ``Fidelity Susceptibility as Holographic PV-Criticality", Phys. Lett.{\bf B} {\bf 765} (2017) 154-158, [arXiv:1604.06909].

\bibitem{Holographic complexity4}
Davood Momeni, Seyed Ali Hosseini Mansoori, Ratbay Myrzakulov, ``Holographic Complexity in Gauge/String Superconductors", Phys. Lett.{\bf B} {\bf 756} (2016) 354-357, [arXiv:1601.03011].

\bibitem{BES}
Souvik Banerjee, Johanna Erdmenger, Debajyoti Sarkar, ``Connecting Fisher information to bulk entanglement in holography", [arXiv:1701.02319].

\bibitem{bak}
D. Bak, ``Information metric and Euclidean Janus correspondence," Phys. Lett. {\bf B} {\bf 756} (2016) 200-204 	[arXiv:1512.04735 [hep-th]].

\bibitem{zhq}
H.-Q. Zhou, ``Renormalization group flows and quantum phase transitions: fidelity versus entanglement'', [arXiv:0704.2945].

\bibitem{CHM}
H. Casini, M. Huerta, and R. C. Myers, ``Towards a derivation of holographic entanglement entropy", JHEP {\bf1105} (2011) 036 [arXiv:1102.0440 [hep-th]].

\bibitem{Janus1}
Dongsu Bak, Michael Gutperle, Shinji Hirano, ``A Dilatonic deformation of AdS(5) and its field theory dual", JHEP {\bf 0305} (2003) 072 [hep-th/0304129].

\bibitem{Janus2}
Dongsu Bak, Michael Gutperle, Shinji Hirano, ``Three dimensional Janus and time-dependent black holes", JHEP {\bf 0702} (2007) 068 [hep-th/0701108].


\bibitem{hhtz}
P. Hauke, M. Heyl, L. Tagliacozzo, P. Zoller, ``Measuring multipartite entanglement via dynamic susceptibilities'', Nature Physics  {\bf 12}, 778-782 , [arXiv:1509.01739 [quant-ph]].

\bibitem{KGP}
M. Kolodrubetz, V. Gritsev, and A. Polkovnikov, ``Classifying and measuring geometry of a quantum ground state manifold", Phys. Rev. {\bf B} {\bf 88} (2013) 064304 [arXiv:1305.0568 [cond-mat.stat-mech]].

\bibitem{jzhang}
J. Zhang, X. Peng, N. Rajendran, and D. Suter, ``Detection of Quantum Critical Points by a Probe Qubit", Phys. Rev. Lett. {\bf 100} (2008) 100501 [arXiv:0709.3273 [quant-ph]].

\bibitem{islam}
R. Islam, R. Ma, P. M. Preiss, M. E. Tai, A. Lukin, M. Rispoli, and M. Greiner ``Measuring entanglement entropy through the interference of quantum many-body twins'', [arXiv:1509.01160].

\bibitem{error}
A. Almheiri, X. Dong and D. Harlow, ``Bulk locality and quantum error correction in AdS/CFT", JHEP {\bf 1504} (2015) 163 [arXiv:1411.7041].

\bibitem{error2}
F. Pastawski, B. Yoshida, D. Harlow, and J. Preskill, ``Holographic quantum error-correcting codes: Toy models for the bulk/boundary correspondence", JHEP {\bf 1506} (2015) 149 [arXiv:1503.06237].

\bibitem{vidal}
H-Q Zhou, R. Or\'{u}s and G. Vidal, ``Ground State Fidelity from Tensor Network Representations," Phys. Rev. Lett. {\bf100} (2008) 080601 [arXiv:0709.4596[cond-mat.stat-mech]].
\end{thebibliography}
\end{document}